\documentclass[10pt,twocolumn]{IEEEtran}

\usepackage{amssymb}
\usepackage{amsmath}
\usepackage{graphicx}
\usepackage[latin1]{inputenc}
\usepackage{graphicx}
\usepackage{cite}
\usepackage{float}
\usepackage{url}
\usepackage{color}

\newtheorem{proposition}{Proposition}
\newtheorem{remark}{Remark}

\newtheorem{lemma}{Lemma}
\newtheorem{corollary}{Corollary}

\begin{document}

\title{Throughput Maximization in Multi-Hop Wireless Networks under Secrecy Constraint}
\author{
\IEEEauthorblockN{Pedro. H. J. Nardelli~\IEEEmembership{Member, IEEE}, Hirley Alves~\IEEEmembership{Student Member, IEEE},\\ Carlos H.~M. de~Lima~\IEEEmembership{Member, IEEE} and Matti Latva-aho~\IEEEmembership{Senior Member, IEEE}} 
\thanks{H. Alves, P. H. J. Nardelli and M. Latva-aho are with the Centre for Wireless Communications (CWC), University of Oulu, Finland. E-mail:\{halves, nardelli, matla\}@ee.oulu.fi.
C. H. M. de Lima is also with São Paulo State University (UNESP), São João da Boa Vista, Brazil. Email: carlos.lima@sjbv.unesp.br 
The authors also would like to thank Aka and Infotech Oulu Graduate School from Finland, and CNPq 490235/2012-3 and Special Visiting Researcher fellowship CAPES 076/2012 from Brazil.}}

\maketitle

\begin{abstract} 
This paper analyzes the achievable throughput of multi-hop sensor networks
for industrial applications under secrecy constraint.
The evaluation scenario comprises
sensors that measure some relevant information of the plant that is first
processed by an aggregator node and then sent to the control unit.
To reach the control unit, a message may travel through relay
nodes, which form a multi-hop
wireless  link.
At every hop, eavesdropper nodes attempt to acquire the messages transmitted through the legitimate link.
The communication design problem posed here is how to maximize the multi-hop throughput from the aggregator to the control unit by finding the best combination of relay positions (i.e. hop length: short or long) and coding rates (i.e. high or low spectral efficiency) so that the secrecy constraint is satisfied.
Using a stochastic-geometry formulation, we show that the
optimal choice of coding rate is normally high and depends on
the path-loss exponent only, while a greater
number of shorter hops are preferable to smaller number of longer hops in
any situation.
For the investigated scenarios, we prove that the optimal
throughput subject to the secrecy constraint achieves the unconstrained
optimal performance -- if a feasible solution exists.
\end{abstract}


\section{Introduction}
Industrial environment imposes challenging conditions on radio propagation due to their
commonplace reflective and absorbent surfaces, as well as electromagnetic interference from the machinery \cite{StenumgaardCM2013}. 
Recently, wireless solutions for industrial applications have gained considerable attention from both academia and industry, using the concept of machine-to-machine communications \cite{Stojmenovic2014,ART:GOH-TIIF15, ART:RAJANDEKAR-IOTJ15, ART:OSSEIRAN-CM14, ART:BOCCARDI-CM14}. 
Such idea enables seamless exchange of information between autonomous devices without any (direct) human intervention. Another advantage of wireless machine-to-machine communications is its scalability, which reduces deployment and maintenance costs. 

In industrial plants, exchange of information is often needed among the
machinery, monitoring devices and control unit; thereby, reliability, low
latency and security become major concerns in the communication system design
\cite{GungorTIE2009}. 
In this context, multi-hop machine-to-machine communications appear as a
promising technology to tackle the industrial environment challenges.
As pointed out in \cite{ART:GOH-TIIF15}, multi-hop schemes are more suitable in such environments with additional interference.

In a typical plant, the design of a multi-hop link between the aggregator and
the control unit can be simplified by setting two parameters:
position of relay nodes and coding rate (spectral efficiency).
The most straightforward design option would be to
use long-hops (less use of network resources) and to set high coding
rates (i.e. more efficient messages in bits/s/Hz).

Industrial networks usually employ unlicensed frequency bands and consequently
are exposed to stronger co-channel interference.
If this is the case, using long
hops in conjunction with high rates may not be the best choice as far as
the former leads to lower signal-to-interference ratio
(SIR) while the latter leads to higher SIR
thresholds needed
to successfully decode a message \cite{Nardelli2012}.

In large industrial deployments, there are various sensors and machines continuously monitoring several
processes.
The resulting information that needs to be exchanged is frequently
confidential, what requires the communication to be reliable, efficient, and secure at all levels of the network infrastructure \cite{GungorTIE2009,
ShiuPHYTutoralWC2011}.
Due to the broadcast nature of the wireless medium, non-intended nodes --
commonly named eavesdroppers -- within the communication range of a given
transmitter can overhear the so-called legitimate
transmission and possibly extract private information
\cite{ShiuPHYTutoralWC2011}.
To avoid that, cryptographic techniques are usually
implemented in the higher layers of the communication protocols to ensure
confidentiality \cite{BOOK:KAUFMAN2002}.

Such techniques, however, depend on secret keys and rely on the  limited
computational power of eavesdroppers, as well as the
reliability guaranteed by channel coding at the physical-layer design. 
These assumptions may not always hold since devices with high computational
power are getting cheaper and widespread. 
Moreover, they become expensive
and difficult to achieve as the network scales \cite{ShiuPHYTutoralWC2011,
MukherjeeCSurveyTut2014}.
In this context, physical-layer security comes as a promising
alternative to complement cryptographic solutions, by adding not only
security  at the physical-layer with strategies that guarantee
reliability, but also confidentiality regardless of
eavesdroppers computational power \cite{ShiuPHYTutoralWC2011, MukherjeeCSurveyTut2014}.

Another interesting solution when dealing with wireless communication over multiple hops is the well-known cooperative relaying strategies \cite{GomezCubaCST2012, MansourkiaieCST2015}. 
As pointed out in \cite{GomezCubaCST2012}, such schemes are robust to
fading and interference impairments due to the enhanced diversity.
Additionally, as discussed in \cite{ZouNetwork2015, J_AlvesSPL2015a}, cooperative diversity schemes also enhance the performance of networks secured at the physical-layer. 

\begin{figure}[!t]
	\centering
	\includegraphics[width=0.75\columnwidth]{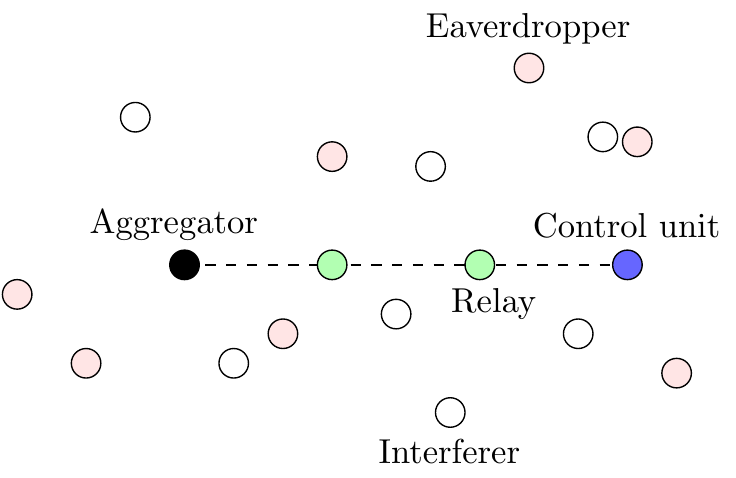}
	%
	\caption{Schematic example of the proposed scenario. The black node is the aggregator (source), the blue node is the control unit (destination) and the green nodes are the relays, all of them defining the legitimate link. The white nodes are the interferers while the red nodes are the eavesdroppers, which attempt to illegitimately acquire the messages send through the multi-hop link. The network designer aims at maximizing the multi-hop throughput by properly deploying the relays and setting the coding rate used while respecting a given secrecy constraint.}
	\label{fig-scheme}
	%
\end{figure}

All in all, the existence of multiple hops, interferers and eavesdroppers 
further complicate the design of wireless communication systems in industrial applications.
Fig.~\ref{fig-scheme} exemplifies an industrial deployment, where several sensors communicate to an aggregator (black node), which by its turn communicates via relays with the control unit (blue node). For instance, an aggregator can act as a relay and help to convey the information to the control unit.
The legitimate link is composed by an aggregator (black node), relays (green nodes) and the control unit (blue node). 
All other randomly distributed nodes in the network are assumed to be either interferers (white nodes) or (potential) eavesdroppers (red nodes). 

We assume that sensors are scattered throughout the industrial
facility to measure  relevant information, which is 
processed by an aggregator node and then sent to the control unit. 
Note that the sensor measurements, their communication with the
aggregetor and the information processing are all assumed to be perfect.
To reach the control unit, the message may travel through relay nodes, forming a multi-hop, wireless link.
At every hop, eavesdropper nodes attempt to acquire the messages transmitted through the legitimate link.

For instance, the aggregator could attempt a single
transmission via long hops, which means that
the channel is used less times and then there is a lower chance of the message being decoded by the eavesdropper.
At the same time this increases the chance that an eavesdropper, which is
closer to the transmitter than the desired receiver, intercepts and
acquires the information being transmitted. 

As we can observe, there are trade-offs regarding possible  eavesdropper locations, number of hops, transmit power and decoding
capabilities, which are function of the interference level perceived at
the receivers. 
To assess such trade-offs, we introduce a tractable model based on to characterize the uncertainty related to interferers (jammers) and eavesdroppers positions and then proceed with a throughput optimization subject to a secrecy constraint.
Then, the main contributions of this paper can be summarized as follows: 
\vspace{1ex}
\begin{itemize}
	\item Analysis of the throughput of industrial communication networks under a secrecy constraint by employing a model that characterizes the uncertainty related to interferers' and eavesdroppers' positions.
	\vspace{1ex}
	\item Closed-form solutions for the optimal multi-hop throughput considering or not the secrecy constraint as a function of network deployments parameters.
	\vspace{1ex} 
	\item Identification of the operational regions proving that the optimal throughput subject to the secrecy constraint achieves the optimal performance if a feasible solution exists.
	\vspace{1ex}
\end{itemize}

The remainder of this paper is organized as follows: Section~\ref{sec-sys-model} introduces the system model and the main metrics used to evaluate the performance of the network.
Section~\ref{sec_unconst_opt} and Section~\ref{sec_const_opt} evaluate the trade-offs involved in the design and deployment of the network. Both sections offer comprehensive numerical results and discussions. Next, Section~\ref{implementation} contextualizes our results and current industrial standards.
Finally, in Section~\ref{conclusions} conclusions and final remarks are drawn.

\section{System model}
\label{sec-sys-model}
Let $D>0$ be the distance from the aggregator to the central unit, assuming
that there exist in-between relay nodes employing a
decode-and-forward strategy \cite{GomezCubaCST2012}.
We consider that the relay nodes are deployed in the straight line defined by the aggregator and the central unit such that the distance $d>0$ between any two nodes is the same.
The number of hops is then computed as $h=D/d$.
We assume that randomly scattered nodes attempt to jam the communication between the aggregator and the central unit. Then, if we assume that the jamming signals experienced by each receiver node along the multi-hop link are
independent, the respective throughput $\mathcal{T}$, with respect to the multi-hop link, can be computed as
\cite{Nardelli2012}:
\begin{equation}
\label{eq-multihop-1}
	\mathcal{T} = \dfrac{\log(1 + \beta)}{h} \; (P_\mathrm{suc})^h,
\end{equation}
where $P_\mathrm{suc}$ is the probability that the message is successfully decoded by the receiver and $\beta>0$ is the minimum required SIR for a successful reception. 
If we assume point-to-point Gaussian codes and interference-as-noise decoding rule \cite{Baccelli2011}, the spectral efficiency of $\log(1 + \beta)$, measured in bits/s/Hz, in the single-hop links is achievable if SIR$>\beta$.

If the network designer choose one long hop $h=1$, the throughput is $\log(1 + \beta) \; P_\mathrm{suc}$.
If more hops are desired, then more network resources are required and the
overall spectral efficiency decreases in relation to the number of hops (i.e. the same information is transmitted at the expense of more channel uses). 
Nevertheless, if more hops are added, $P_\mathrm{suc}$ is expected to increase.
These contradictory effects are captured by \eqref{eq-multihop-1}.

We assume a field of jammers (malicious interferers) that is characterized by a 2-dimension uniform Poisson point process $\Phi_\textup{int}$  with intensity $\lambda_\textup{int}>0$, measured in transmitters per unit of area \cite{haenggi2012stochastic}.
We assume that the channel has two components: one related to the distance-dependent path-loss such that the received power decays with the distance and other related to fading.
The received power at the node of interested can be computed as $g_i
r_i^{-\alpha}$, where $r_i$ is the distance between the reference receiver and
the $i\text{th}$ node, $g_i$ is the channel gain between them,
and $\alpha>2$ the path-loss exponent \cite{Inaltekin2009}.

We consider the communication occurs in time-slot basis so that the slot length is the time required to transmit one packet \cite{Nardelli2013a}.
If the nodes' positions and the channel gains do not change during the packet transmission, the signal-to-interference ratio (SIR) is computed as\footnote{We assume here interference-limited networks. As pointed in \cite{Weber2010}, the inclusion of the noise power leads to a more complex analysis without providing any significant qualitative difference.}
\begin{equation} 
\label{eq_SIR}
	\textrm{SIR} = \dfrac{g_{0} d^{-\alpha}}{\underset{i \in \Phi_\textup{int}}{\displaystyle \sum} g_{i} r_i^{-\alpha}}.
\end{equation}

To compute $P_\mathrm{suc}$, we assume that the channel gains $g_i$ are independent and identically distributed exponential random variables (Rayleigh fading) and that the interferers' positions change every time slot.
In this way, each time-slot is a different realization of the point processes $\Phi_\textup{int}$ and the channel gains $g_i$.
From this assumptions, the success probability is \cite{haenggi2012stochastic}:
\begin{equation} 
\label{eq_Psuc}
	P_\mathrm{suc}= e^{-  \lambda_\mathrm{int} \kappa \pi d^2 \beta^{2/\alpha}},
\end{equation}
where $\kappa = \Gamma(1 + 2/\alpha)\Gamma(1 - 2/\alpha)$.

Let us now consider that there are eavesdroppers that are capable of
decoding the transmitted messages if the SIR experienced by them are greater than the threshold $\beta_\mathrm{eav}>0$.\footnote{This threshold may reflect how powerful are the eavesdroppers: a low $\beta_\mathrm{eav}$
indicates that the eavesdroppers are able to
successfully decode messages even with low SIR, reflecting
a powerful decoding scheme.}
Their spatial distribution are modeled as a 2-dimensional uniform Poisson point process $\Phi_\mathrm{eav}$ with intensity $\lambda_\mathrm{eav}>0$, measured in eavesdroppers per unit of area.
The channel gains in relation to the transmitter are modeled as in the interferers' process described above (quasi-static Rayleigh fading and distance-dependent path-loss).
As before, a different realization of the point process and channel gains are assumed at every different time-slot.
In this scenario, due to the lack of any side information regarding the specific position and the channel gains, it is not possible to guarantee 100\% of secrecy in the communication of the desired link.
%

Herein, we assume a secrecy constraint referring to the aggregator--control unit multi-hop link.
In this case, the probability that the eavesdropper illegitimately acquires the message should be, statistically, lower than or equal to $\epsilon$\%.
To compute such probability, we need to evaluate the outage probability in the
eavesdropper\footnote{In fact this is an approximation since there will be
closer eavesdroppers that
experience a better channel.  This, however, is a good approximation and holds
in most of the cases for the spatial densities considered here.}.
The probability density function $f_{R_1}(r)$ of distance $r$ between an arbitrary point to the closest point of a Poisson point process with intensity $\lambda_\mathrm{eav}$ is given by \cite{haenggi2012stochastic}:
\begin{equation} 
\label{eq_pdf-closest-eav}
	f_{R_1}(r) = \lambda_\mathrm{eav} \;  2 \pi r e^{-  \lambda_\mathrm{eav}  \pi r^2}.
\end{equation}

Similarly to \eqref{eq_Psuc}, the outage probability $P_\mathrm{out:eav}$ (i.e. the probability that the packet is not successfully decoded) at the eavesdropper can be computed as \cite{haenggi2012stochastic}:
\begin{align} 
\label{eq_Pout-eav}
	P_\mathrm{out:eav} &= \mathbb{E}_r \left[1- e^{-  \lambda_\mathrm{int} \kappa \pi r^2  (\beta_\mathrm{eav})^{2/\alpha}}\right] \nonumber \\ \vspace{2ex} 
	&= \dfrac{\lambda_\mathrm{int} \kappa   (\beta_\mathrm{eav})^{2/\alpha}}{\lambda_\mathrm{int} \kappa   (\beta_\mathrm{eav})^{2/\alpha}+\lambda_\mathrm{eav}},
\end{align}
where $\mathbb{E}_r \left[\cdot\right]$ is the expected value in regard to the distance $r$ given by \eqref{eq_pdf-closest-eav}.

%
We are now ready to state the optimization problem of interest as follows: \textit{what are the hop length $d$ and SIR threshold $\beta$ that jointly optimize the multi-hop throughput $\mathcal{T}$ given by \eqref{eq-multihop-1} while the secrecy constraint $\epsilon$ is satisfied?}
Mathematically, we have the following:
\begin{equation}
\label{eq-opt}
	\begin{aligned}
		&  \underset{(\beta, d)}{\max} & & \mathcal{T} = \dfrac{d \log(1 + \beta)}{D} \; \left(e^{-  \lambda_\mathrm{int} \kappa \pi d^2 \beta^{2/\alpha}}\right)^{D/d} \\
		&   \ \text{s.t. } 	& & \left(\dfrac{\lambda_\mathrm{int} \kappa   (\beta_\mathrm{eav})^{2/\alpha}}{\lambda_\mathrm{int} \kappa   (\beta_\mathrm{eav})^{2/\alpha}+\lambda_\mathrm{eav}}\right)^{D/d} \geq 1- \epsilon
	\end{aligned},
	\vspace{2ex}
\end{equation}
where the constraint is the probability that the eavesdropper links are in outage at every hop of the multi-hop link with a probability greater than equal to $1 - \epsilon$.

\section{Unconstrained optimization} \label{sec_unconst_opt}
Let us start presenting the solution of the unconstrained optimization problem assuming that the number of hops $h$ can be a real number so that the hop length $d$ can assume any positive value for a given $D$.

\begin{proposition}
\label{prop-unconst-opt}
The pair $(\beta^*, d^*)$ of the unconstrained version of the optimization problem given by \eqref{eq-opt} is:
\begin{eqnarray}
\label{eq-opt-unconst-setup-beta}
	\beta^*	& = -1 + e^{\mathcal{W}_0\left(-\frac{\alpha}{2}e^{-\alpha/2}\right)+\frac{\alpha}{2}}  \\ 
\label{eq-opt-unconst-setup-d}
	d^*	& =  \dfrac{1}{D \lambda_\mathrm{int} \kappa \pi (\beta^*)^{ 2/\alpha} },
\end{eqnarray}
where $\mathcal{W}_0(\cdot)$ is the principal branch of the Lambert W function\footnote{We have used the function $\mathrm{LambertW}(\cdot)$ from the library SymPy \cite{lambertw}.} \cite{Nardelli2013a}, which is defined as $x=\mathcal{W}_0(x) e^{\mathcal{W}_0(x)}$ such that $x\geq - e^{-1}$ and $\mathcal{W}_0(x) \geq -1$.

The optimal throughput $\mathcal{T}^*$ is then:
\begin{equation}
\label{eq-opt-uncos}
	\mathcal{T}^* = \dfrac{\log\left( e^{\mathcal{W}_0\left(-\frac{\alpha}{2}e^{-\alpha/2}\right)+\frac{\alpha}{2}} \right)}{e \log(2) D \lambda_\mathrm{int} \kappa \pi \left(-1 + e^{\mathcal{W}_0\left(-\frac{\alpha}{2}e^{-\alpha/2}\right)+\frac{\alpha}{2}}\right)}.
\end{equation}
\end{proposition}

\begin{IEEEproof}[Outline of proof]
The first step is to show that multi-hop throughput
$\mathcal{T}$ in \eqref{eq-opt} is a quasi-concave function in terms of
both $d$ and $\beta$.
Then, the pair $(\beta^*, d^*)$  that leads to the optimal unconstrained
throughput $\mathcal{T}^*$ can be found as the joint solution of the
following partial derivative equations $\partial \mathcal{T}/ \partial d
= 0$ and $\partial \mathcal{T}/ \partial \beta = 0$.

Solving that system of equation, we find the equilibrium point $(\beta^*, d^*)$  that is given by \eqref{eq-opt-unconst-setup-beta} and \eqref{eq-opt-unconst-setup-d}.
Inserting these values into multi-hop throughput given by \eqref{eq-opt}, we obtain equation \eqref{eq-opt-uncos}.
\end{IEEEproof}

\begin{figure}[!t]
	\includegraphics[width=\columnwidth]{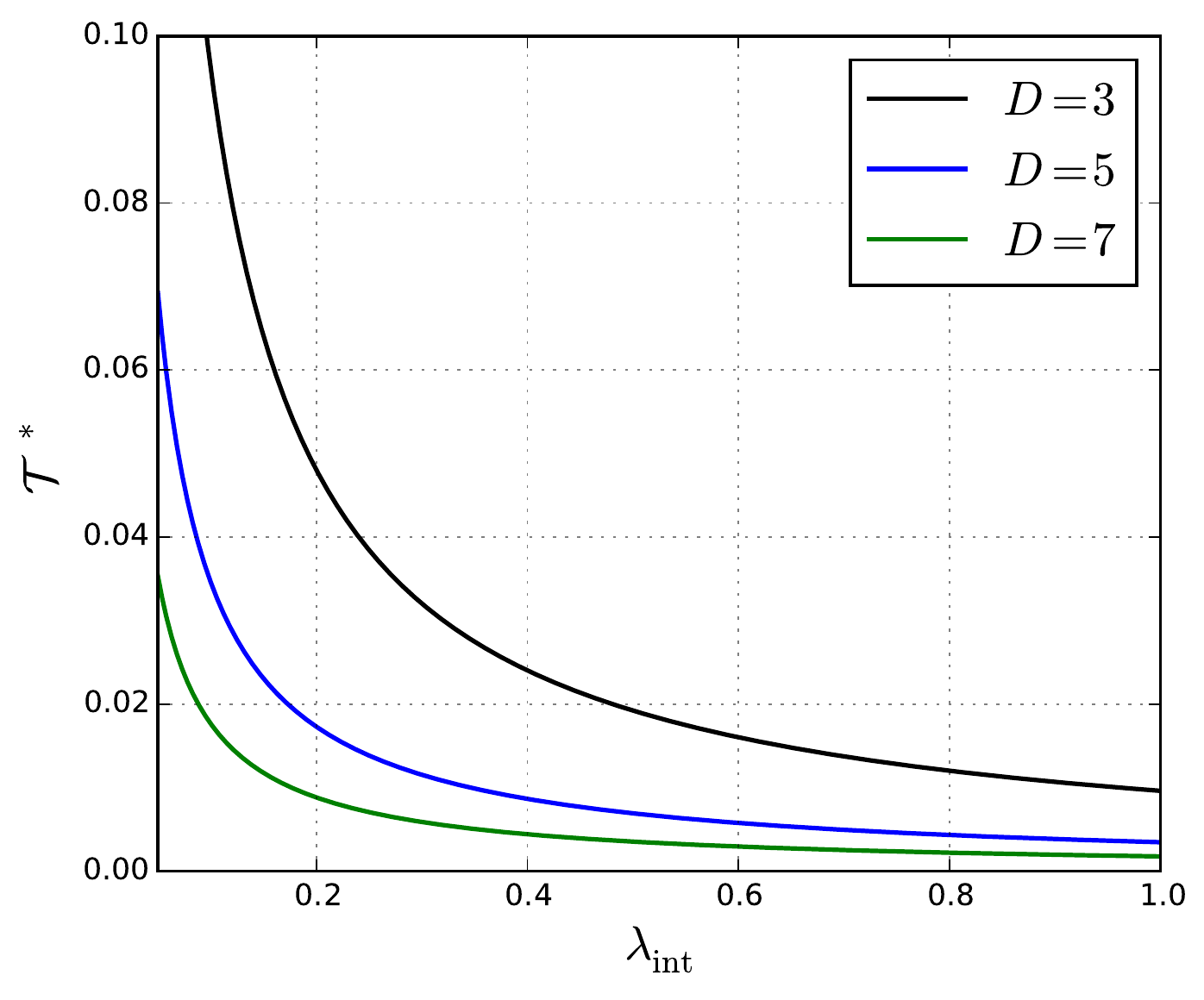}
	\vspace{-0.8cm}
	\caption{Optimal multi-hop throughput $\mathcal{T}^*$, given by \eqref{eq-opt-uncos}, as a function of the density of interferers $\lambda_\mathrm{int}$ for different values of multi-hop distance $D$, considering $\alpha = 4$.}
	\label{fig-T-vs-Lint-uncons}
\end{figure}

Fig. \ref{fig-T-vs-Lint-uncons} exemplifies how the optimal throughput $\mathcal{T}^*$ varies with the density of interferers  $\lambda_\mathrm{int}$ for different multi-hop distances $D$.
One can see that the lower densities $\lambda_\mathrm{int}$ yields higher optimal throughputs, regardless of the multi-hop distance considered.
Looking at the multi-hop distances, we find that the lower the distance $D$, the higher the throughput $\mathcal{T}^*$.

Although those results are somehow expected, it is interesting to analyze the reasons behind this behavior, which will later help us
to understand the solution of the constrained optimization.
From \eqref{eq-opt-unconst-setup-beta}, the optimal value of
$\beta^*$ is independent of any other parameter of the system, but the
path-loss exponent $\alpha$, which is not under the designer control.
Therefore, $\beta^*$ is fixed if $\alpha$ is fixed and the single-hop distance $d^*$ is the variable that changes with $\lambda_\mathrm{int}$ and/or $D$, as indicated by \eqref{eq-opt-unconst-setup-d}.

Equation \eqref{eq-opt-uncos} shows that the optimal throughput $\mathcal{T}^*$ is inversely proportional to $\lambda_\mathrm{int}$ and $D$.
It is worth noting that for small values of $D$ and/or $\lambda_\mathrm{int}$, $\mathcal{T}^*$ tends to infinity. 
This is a byproduct of our assumptions and clearly does
not represent actual scenarios. 
Although, we understand this limitation, we still believe that the simplicity of our results can provide clear, and reasonable, guidelines on the network design.

When the same $\lambda_\mathrm{int}$ is considered, the optimal throughput $\mathcal{T}^*$ is determined only by $D$: if a packet needs to travel longer source-destination distances, the single-hops should be surprisingly smaller to support the optimal coding rate $\beta^*$.
This happens because the smaller the single-hop distance, the higher the $\mathrm{SIR}$ experienced by the receiver nodes.
In this case, having more shorter hops is statistically more advantageous  than having less longer hops.
A similar analysis is also valid when assessing the case when the same multi-hop distance is assumed $D$ and the intensity $\lambda_\mathrm{int}$ is varying.

Although the results showing that longer distances $D$ and
higher intensity of interferer nodes $\lambda_\mathrm{int}$ degrade the
throughput are expected, the design choices $(\beta^*, d^*)$ that optimize the
multi-hop throughput $\mathcal{T}$ are rather surprising.
In the next section, we will see how the secrecy constraint will affect the optimal system design.

\section{Constrained optimization} \label{sec_const_opt}
\label{subsubsec-constrained}
Let us now focus on the optimization problem subject to the secrecy constraint stated in \eqref{eq-opt}.
We first recall that the network designer does not have any control on the eavesdropper parameters so that $\lambda_\mathrm{eav}$ and $\beta_\mathrm{eav}$ are input variables (i.e. external factors).

\begin{lemma}
\label{lemma-const-opt}
The secrecy constraint given by \eqref{eq-opt} can be rewritten as

\begin{equation}
\label{eq-newcontraint}
	d \leq d_\mathrm{c} \leq D,
\end{equation}
where the new constraint $d_\mathrm{c}$ is given by:
\begin{equation}
\label{eq-newcontraint-2}
	d_\mathrm{c} = \dfrac{D \; \log\left(\frac{\lambda_\mathrm{int} \kappa   (\beta_\mathrm{eav})^{2/\alpha}}{\lambda_\mathrm{int} \kappa   (\beta_\mathrm{eav})^{2/\alpha}+\lambda_\mathrm{eav}}\right)}{\log(1-\epsilon)}.
\end{equation}
\end{lemma}

\begin{IEEEproof}
We start by manipulating the secrecy constraint from \eqref{eq-opt} as follows:
\begin{eqnarray}
 &\dfrac{D}{d} \log \left(\dfrac{\lambda_\mathrm{int} \kappa   (\beta_\mathrm{eav})^{2/\alpha}}{\lambda_\mathrm{int} \kappa   (\beta_\mathrm{eav})^{2/\alpha}+\lambda_\mathrm{eav}}\right) \geq \log(1- \epsilon) & \Rightarrow \nonumber  \\ [0.4cm]
 \label{eq-ineq-dist-1}
  \Rightarrow & \dfrac{D \; \log \left(\dfrac{\lambda_\mathrm{int} \kappa   (\beta_\mathrm{eav})^{2/\alpha}}{\lambda_\mathrm{int} \kappa   (\beta_\mathrm{eav})^{2/\alpha}+\lambda_\mathrm{eav}}\right)}{\log(1- \epsilon)}   \geq \; d. &
\end{eqnarray}

We now use the fact that the single-hop must have a length lower than or equal to the multi-hop $d\leq D$ and that  the distances are strictly positive to conclude this proof.
\end{IEEEproof}

\vspace{1ex}

\begin{proposition}
\label{prop-const-opt}
The solution of the constrained optimization problem stated in \eqref{eq-opt} is given in Proposition \ref{prop-unconst-opt} with $\log \left(\frac{\lambda_\mathrm{int} \kappa   (\beta_\mathrm{eav})^{2/\alpha}}{\lambda_\mathrm{int} \kappa   (\beta_\mathrm{eav})^{2/\alpha}+\lambda_\mathrm{eav}}\right) \leq \log(1- \epsilon) \leq D^2 \lambda_\mathrm{int} \kappa \pi  (\beta_\mathrm{eav})^{2/\alpha} \log \left(\frac{\lambda_\mathrm{int} \kappa   (\beta_\mathrm{eav})^{2/\alpha}}{\lambda_\mathrm{int} \kappa   (\beta_\mathrm{eav})^{2/\alpha}+\lambda_\mathrm{eav}}\right)$.
\end{proposition}
\vspace{1ex}

\begin{IEEEproof}
	Let us start by considering the second part of the
	secrecy constraint given by Lemma \ref{lemma-const-opt}, namely
	$d_\mathrm{c} \leq D$ .
If $\log \left(\frac{\lambda_\mathrm{int} \kappa   (\beta_\mathrm{eav})^{2/\alpha}}{\lambda_\mathrm{int} \kappa   (\beta_\mathrm{eav})^{2/\alpha}+\lambda_\mathrm{eav}}\right) > \log(1- \epsilon)$, then $d_\mathrm{c} > D$, the secrecy constraint is then violated and the problem has no feasible solution.

In the case where $\log \left(\frac{\lambda_\mathrm{int} \kappa
(\beta_\mathrm{eav})^{2/\alpha}}{\lambda_\mathrm{int} \kappa
(\beta_\mathrm{eav})^{2/\alpha}+\lambda_\mathrm{eav}}\right) \leq \log(1-
\epsilon)$, we need to verify the constraint $d \leq
d_\mathrm{c}$.
As stated in Proposition \ref{prop-unconst-opt}, the optimal choice of $\beta^*$ only depends on $\alpha$.
Therefore we focus on the optimal single-hop distance $d^*$ given by
\eqref{eq-opt-unconst-setup-d}: the optimal solution can be only
obtained if the inequality $d^* = \dfrac{1}{D
\lambda_\mathrm{int} \kappa \pi (\beta^*)^{ 2/\alpha} } \leq d_\mathrm{c}$
is satisfied.  
\end{IEEEproof}

\vspace{1ex}

\begin{corollary}
\label{cor-opt}
The solution of the constrained optimization problem stated in \eqref{eq-opt} does not exist if  $\log\left(\frac{\lambda_\mathrm{int} \kappa   (\beta_\mathrm{eav})^{2/\alpha}}{\lambda_\mathrm{int} \kappa   (\beta_\mathrm{eav})^{2/\alpha}+\lambda_\mathrm{eav}}\right) > \log(1- \epsilon)$ and then $\mathcal{T}^* = 0$.
\end{corollary}

\vspace{1ex}

\begin{IEEEproof}
	This proof follows from the first part of the proof of Proposition
	\ref{prop-const-opt}, when $\log \left(\frac{\lambda_\mathrm{int} \kappa
	(\beta_\mathrm{eav})^{2/\alpha}}{\lambda_\mathrm{int} \kappa
	(\beta_\mathrm{eav})^{2/\alpha}+\lambda_\mathrm{eav}}\right) > \log(1-
	\epsilon)$ implies that the problem has no feasible solution.
\end{IEEEproof}

\vspace{1ex}

\begin{remark}
	In the scenarios under investigation, the inequality $d^* \leq d_\mathrm{c}$ holds due to the
	combination of the system parameters and target variables.
\end{remark} 
\vspace{1ex}

From this remark and the analytic results previously stated, the solution of optimization problem with secrecy constraint only depends on the relation between $d_\mathrm{c}$ and the multi-hop distance $D$ for the cases studied here.
More specifically, Corollary \ref{cor-opt} tells us that the optimal solution
exists if the secrecy constraint $\epsilon$  is achievable for the network
density of interferers $\lambda_\mathrm{int}$, density of
eavesdroppers $\lambda_\mathrm{eav}$ and their SIR threshold
$\beta_\mathrm{eav}$ considered.

\begin{figure*}[!t]
	\includegraphics[width=\textwidth]{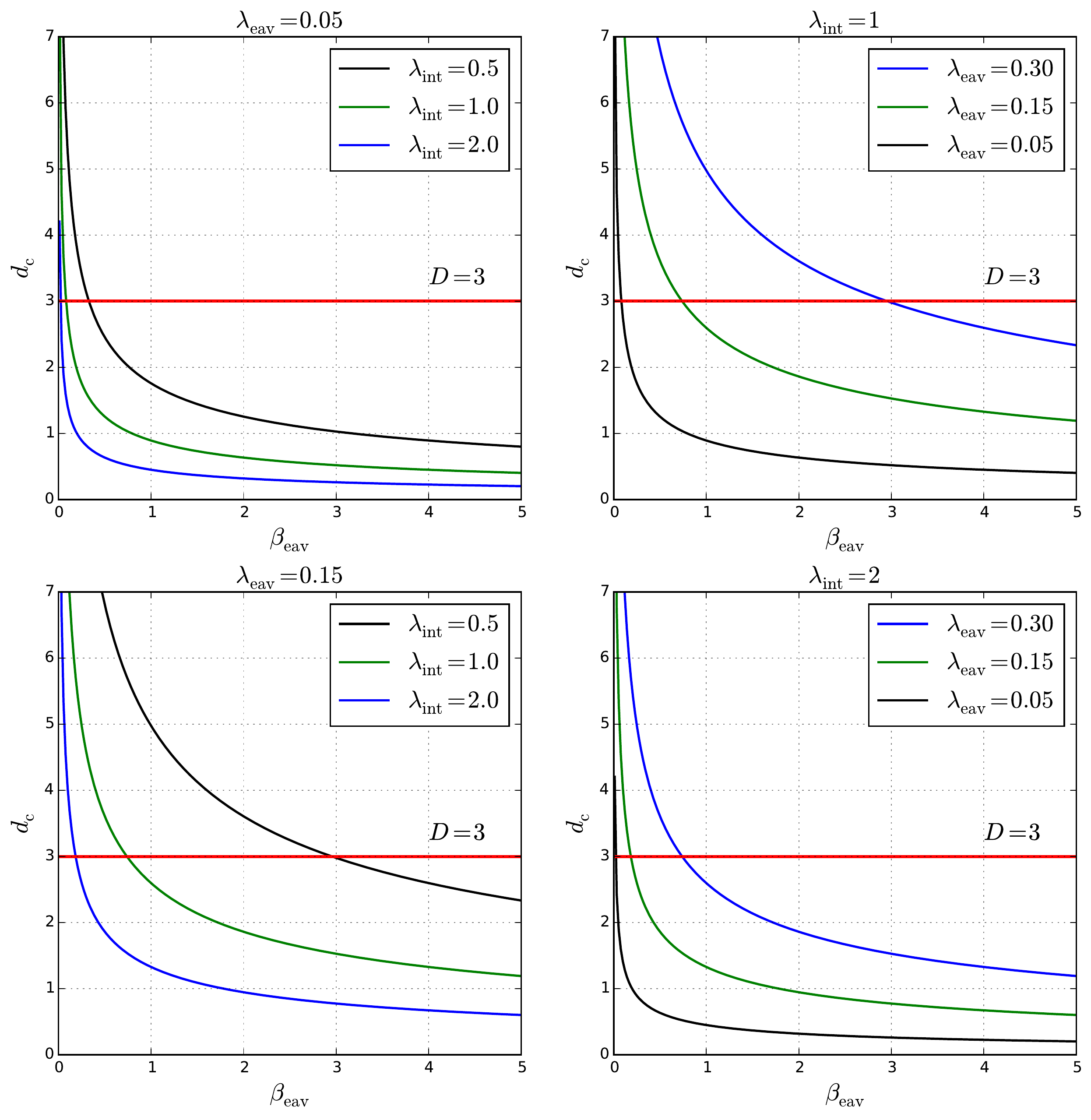}
	%
	\caption{Distance constraint $d_\mathrm{c}$ as a function of the eavesdroppers' SIR threshold $\beta_\mathrm{eav}$ for different combination of densities $\lambda_\mathrm{int}$ and $\lambda_\mathrm{eav}$, assuming the path-loss exponent $\alpha=4$ and secrecy constraint $\epsilon = 10 \%$. We consider the multi-hop distance $D=3$ that is presented by the red line. When the $d_\mathrm{c}\leq D = 3$, then the optimal solution of \eqref{eq-opt} exists and it is given by Proposition \ref{prop-unconst-opt}.}
	\label{fig-dc-vs-beta-eav}
\end{figure*}

Fig. \ref{fig-dc-vs-beta-eav} (presented in the next page) shows the distance constraint $d_\mathrm{c}$ as a function of the eavesdroppers' SIR threshold $\beta_\mathrm{eav}$ for different combination of densities $\lambda_\mathrm{int}$ and $\lambda_\mathrm{eav}$.
One can see that lower SIR thresholds $\beta_\mathrm{eav}$ cause the unfeasibility of the optimal solution.
As expected, if the eavesdroppers are able to decode messages with low SIR, then their chance of correctly decode the information of the legitimate link grows, regardless of the densities $\lambda_\mathrm{int}$ and $\lambda_\mathrm{eav}$.
The effects of $\lambda_\mathrm{int}$ and $\lambda_\mathrm{eav}$ are the following.
The higher the density of eavesdroppers $\lambda_\mathrm{eav}$, the stricter is the distance constraint $d_\mathrm{c}$.
This is due to the fact that big values of $\lambda_\mathrm{eav}$ lead to greater probabilities that a eavesdropper node is closer to the legitimate link.
This, in turn, requires a more stringent
$d_\mathrm{c}$ to satisfy the secrecy constraint $\epsilon$.
The increase of the density of interferers $\lambda_\mathrm{int}$, on the
other hand, helps the secrecy of the legitimate link.
This is in line with the general literature on
physical layer security (e.g.  \cite{J_AlvesSPL2015b}) since higher
$\lambda_\mathrm{int}$ leads to probabilistically lower SIR.
This will then result in more outages in the eavesdropper links, making the distance constraint $d_\mathrm{c}$ less stringent.

\begin{figure}[!t]
	\includegraphics[width=\columnwidth]{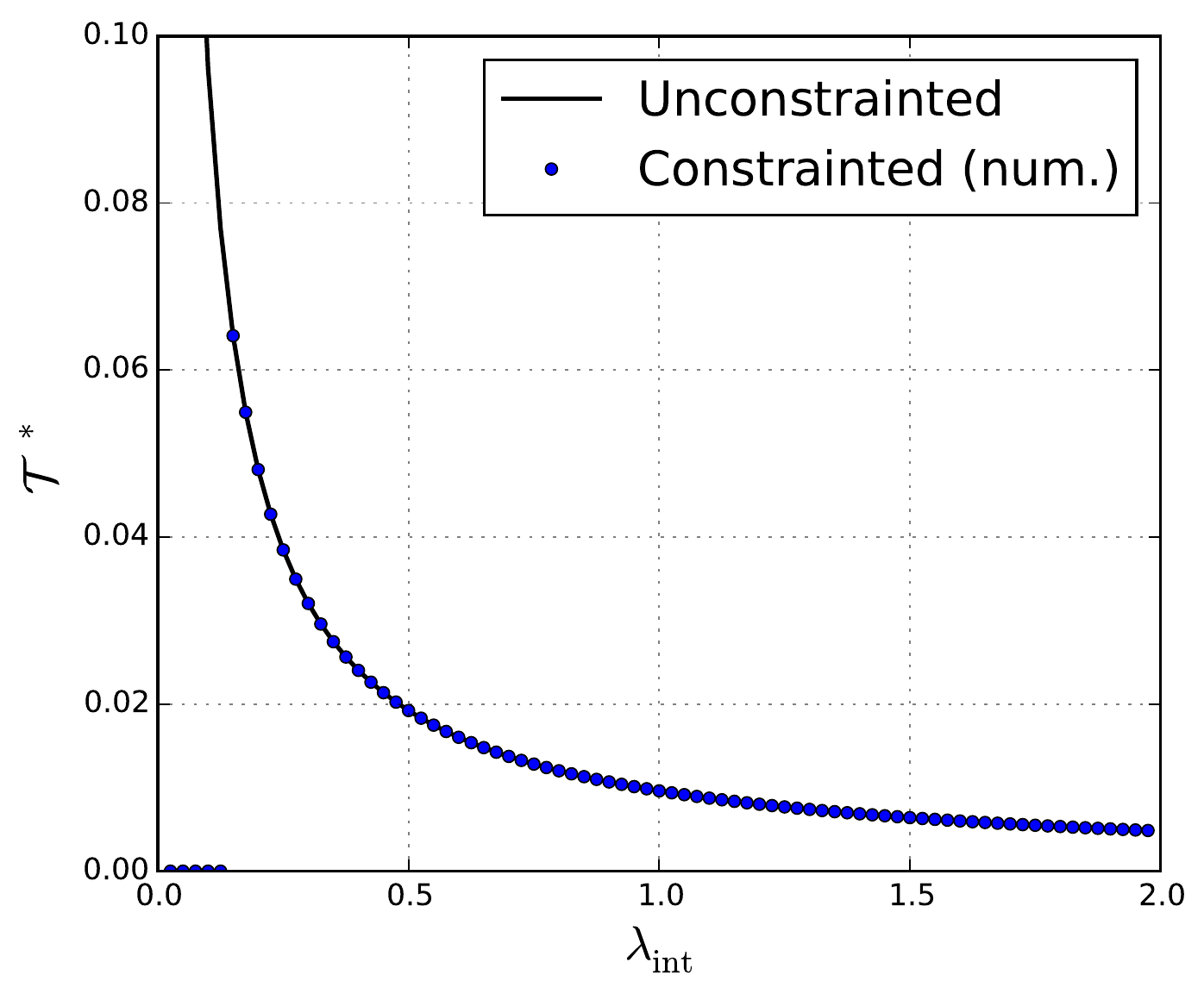}
	\vspace{-0.8cm}
	\caption{Optimal multi-hop throughput $\mathcal{T}^*$  as a function of the density of interferers $\lambda_\mathrm{int}$ for  $D =3$, $\alpha = 4$ and  $\epsilon = 10 \%$. We consider the unconstrained optimization given by \eqref{eq-opt-uncos} and the numerical solution of \eqref{eq-opt}.}
	\label{fig-T-vs-Lint-cons}
\end{figure}
Fig. \ref{fig-T-vs-Lint-cons} shows an example of the optimal constrained multi-hop throughput\footnote{To solve the constrained optimization, we have used the numerical function $\mathrm{fmin\_l\_bfgs\_b}$  from the library SciPy \cite{optimize-scipy}.}, in relation to the unconstrained case.
The optimal multi-hop throughput $\mathcal{T}^*$ is plotted as a function of $\lambda_\mathrm{int}$ for  $D =3$, $\alpha = 4$ and
$\epsilon = 10 \%$.
One can verify that the constrained optimization can achieve the unconstrained
performance if the solution of the problem is within its feasibility region,
which can be analytically determined as predicted in Corollary
\ref{cor-opt}.

\section{Implementation and deployment aspects}\label{implementation}

As previously mentioned, the scenario under analysis is a
simplified, abstract, version of an actual industrial communication network.
In any case, we would like to reinforce the value of our results, which are simple enough to characterize important trade-offs on the system design.
In the proposed model, we do not assume any information about
interferers and (malicious) eavesdroppers, which might be a more
practical consideration.
For example, the design of a physical-layer secured network assumes some
information about eavesdroppers as in
\cite{ShiuPHYTutoralWC2011, MukherjeeCSurveyTut2014}.
Therein, each transmitter attempts to convey its message in a
reliable and confidential way, once they are aware that non-intended receivers
may be overhearing their transmission. 
In the case of interference, there are well-established approaches that uses the channel and/or location information to increase the system throughput.

In what follows, we briefly discuss some of the major standardization efforts,
which could benefit from our results.
For instance, ZigBee (IEEE 802.15.4) and Bluethooth (Low Energy) network standards serve not only for industrial applications, but also for home automation for instance \cite{GungorTIE2009, AlAghaTIE2009}. 
Both standards are low-power and have limited communication range (few tens of
meters in indoor environment), and thus could take advantage of
our guidelines: a large number of hops in short range communication
is preferable over long hops. 
Another possible alternative for wireless industrial is WirelessHART \cite{AlAghaTIE2009}, which has already embedded functionalities that allow information relaying. 

It is worthy noting that our discussions so far are based on industrial environments; however, our results also extend to modern (smart) power grids due to the similarities of the communication environment. For instance, smart grids also present a distinct profile of interference due to the highly reflective materials and electromagnetic interference from the machinery, especially at distribution side. Besides, communication links suffer additional interference from concurrent transmissions from neighboring devices and aggregations as discussed in \cite{ART:Nardelli-AN2015}. 
This initial assessment is then extended in order to include not only reliability analysis but also security and privacy in \cite{AlvesTIFS2015}. 
This evinces the potential applications and relevance of our results. 

\section{Conclusions and Final Remarks} \label{conclusions}

This paper analyzes the throughput of industrial multi-hop machine-to-machine networks under a secrecy constraint. 
The scenario under analysis consists in an aggregator node, which collects and processes the sensor measurements, and a control unit that needs the proceeded information.
This communication is wireless and may occur over multiple hops, and the
communication engineer is expected to find the optimal position of the relay
nodes and the coding rates used in the single-hop links so as
to maximize the throughput in [bits/s/Hz] while respecting a given
secrecy constraint.

By employing our stochastic-geometric-based model to characterize the
uncertainties involved in the eavesdroppers' and
interferers' positions, we first showed that the optimal choice without any
secrecy constraint of coding rate (spectral efficiency) depends only on the
path-loss exponent and  normally assumes high value.
To sustain such high rate, a great number of shorter hops are then preferable to a small number of longer hops. 
When the secrecy constraint is assumed, we proceeded with the throughput optimization and proved that the unconstrained performance can be achieved with the same optimal relay positions and coding rates only if a feasible solution exists.
Otherwise, there is no solution for the problem that satisfies the minimum level of secrecy required.

As a next step, we expect to evaluate our guidelines in actual industrial environments by following the insights provided herein.
To do so, we aim at designing a feasible experimental deployment that utilizes established standards for industrial wireless systems.
It is also important to point out that, although this analysis has been presented focusing on industrial deployments, our framework can be also extended to different kind of smart applications such as homes, cities, energy grids or highways.

\bibliographystyle{IEEEtran}
\bibliography{ref_abbrev}
\end{document}